# Entanglement and impropriety

Brian R. La Cour · Thomas W. Yudichak

**Abstract** The relationship between quantum entanglement and classical impropriety is considered in the context of multi-modal squeezed states of light. Replacing operators with complex Gaussian random variables in the Bogoliubov transformations for squeezed states, we find that the resulting transformed variables are not only correlated but also improper. A simple threshold exceedance model of photon detection is considered and used to demonstrate how the behavior of improper Gaussian random variables can mimic that of entangled photon pairs when coincidence post-selection is performed.

## 1 Introduction

Entanglement is considered a quintessentially quantum property, one with no classical analogue. Schrödinger described it as "*the* characteristic trait of quantum mechanics" [1]. Yet despite its central role in our understanding of quantum physics and its many applications in quantum information science, the fundamental nature of entanglement remains as mysterious as when it was first conceived.

Mathematically, entanglement may be defined as the property of nonseparability for vectors in (or operators on) a tensor product of Hilbert spaces. When combined with the Born rule, this property entails the many observational consequences of quantum entanglement, but as a mathematical property alone it is by no means limited to quantum systems. This trivial observation has given rise to the notion of classical entanglement, wherein the mathematical description of certain classical systems may also be described as nonseparable under a suitable identification of a product Hilbert space (e.g., modes of a vibrating drum) [2–4].

It is important to recognize that mere nonseparability in a classical system does not entail the many curious observational consequences of true quantum entanglement. The physical significance of entanglement lies in the unique statistical characteristics of entangled systems and the nonlocal effects they seem to imply. This behavior has been demonstrated most strikingly in a series of experiments considered to be free of all reasonable loopholes that might permit a local realist interpretation [5–7].

---

B. R. La Cour (✉) · T. W. Yudichak
Applied Research Laboratories, The University of Texas at Austin, P.O. Box 8029, Austin, TX 78713-8029, USA
e-mail: blacour@arlut.utexas.edu

In this paper, we consider an interesting relationship between quantum entanglement and classical statistics that appears, until now, to have gone unnoticed. Inspiration is taken from the notion that many quantum effects may be reproduced by replacing the virtual zero-point field of quantum electrodynamics with one that is real and stochastic [8]. This approach has been used extensively as a method for classical modeling of certain quantum systems [9]. For example, the relationship to entanglement was studied by deriving a Wigner function representation of spontaneous parametric downconversion through a detailed physical modeling of nonlinear optical processes using a classical zero-point field [10,11].

This result may seem surprising since squeezed vacuum states do not admit a positive $P$ representation, even though they have a positive Wigner function. From a mathematical perspective, this is simply a consequence of the optical equivalence theorem, which implies the equivalence of a Gaussian quantum state and a corresponding classical Gaussian random vector [12]. The present work generalizes prior research to arbitrary multi-modal squeezed states arising from symmetric squeezing matrices and examines the relationship to improper complex Gaussian random variables. In addition, the behavior under a deterministic model of photon detection is considered [13–16], which is an approach the previous work had not considered. Although Gaussian states may in some respects be deemed classical, the introduction of a nonlinear measurement scheme, such as we consider, when combined with post-selection can give rise to contextuality and, hence, quantum-like behavior such as violations of the Bell–CHSH inequality [15,17,18].

## 2 Multi-mode squeezing

Let $\boldsymbol{\xi}$ be a $d \times d$ symmetric matrix defining the quantum mechanical multi-mode squeezing operator

$$\hat{S} = \exp\left[\frac{1}{2}\left(\hat{\boldsymbol{a}}^\dagger\right)^\mathsf{T} \boldsymbol{\xi}\, \hat{\boldsymbol{a}}^\dagger - \frac{1}{2}\hat{\boldsymbol{a}}^\mathsf{T} \boldsymbol{\xi}^\mathsf{H} \hat{\boldsymbol{a}}\right], \tag{1}$$

where $(\hat{\boldsymbol{a}}^\dagger)^\mathsf{T} = [\hat{a}_1^\dagger, \ldots, \hat{a}_d^\dagger]$ is a row vector of creation operators over $d$ distinct modes and $\boldsymbol{\xi}^\mathsf{H} = (\boldsymbol{\xi}^*)^\mathsf{T}$ is the Hermitian conjugate of the matrix $\boldsymbol{\xi}$. The latter notation is used to distinguish the conjugate transpose of a matrix from the adjoint of an operator.

We may write $\boldsymbol{\xi}$ in the general polar form $\boldsymbol{\xi} = \mathbf{RQ}$, where $\mathbf{R}$ is positive semi-definite and $\mathbf{Q}$ is unitary. Since $\boldsymbol{\xi}$ is symmetric and, therefore, normal, $\mathbf{R} = (\boldsymbol{\xi}\boldsymbol{\xi}^\mathsf{H})^{1/2}$. If, furthermore, $\mathbf{R}$ is positive definite, then it is also invertible and we may take $\mathbf{Q} = \mathbf{R}^{-1}\boldsymbol{\xi}$. In the degenerate case $\mathbf{R} = \mathbf{0}$, we may simply take $\mathbf{Q} = \mathbf{I}$ to be the identity. More generally, if $\boldsymbol{\xi} = \mathbf{UDV}^\mathsf{H}$ is a singular value decomposition of $\boldsymbol{\xi}$, where $\mathbf{U}$ and $\mathbf{V}$ are unitary and $\mathbf{D}$ is diagonal and positive semi-definite, then $\mathbf{R} = \mathbf{UDU}^\mathsf{H}$ and $\mathbf{Q} = \mathbf{UV}^\mathsf{H}$. Using the polar decomposition $\boldsymbol{\xi} = \mathbf{RQ}$, we may now write the corresponding Bogoliubov transformation of $\hat{\boldsymbol{a}}$, denoted $\hat{\boldsymbol{b}} = \hat{S}^\dagger \hat{\boldsymbol{a}} \hat{S}$, as follows [19]:

$$\hat{\boldsymbol{b}} = (\cosh \mathbf{R})\, \hat{\boldsymbol{a}} + (\sinh \mathbf{R})\, \mathbf{Q}\, \hat{\boldsymbol{a}}^\dagger. \tag{2}$$

A classical analogue will now be considered by replacing $\hat{\boldsymbol{a}}$ with the random vector $\boldsymbol{a} = \sigma \boldsymbol{z}$, where $\boldsymbol{z}$ is a $d$-dimensional standard complex Gaussian random vector representing the $d$ distinct vacuum modes and $\sigma^2 \hbar \omega$ is the modal energy. Specifically, $\boldsymbol{z}$ is a complex Gaussian random vector such that $\mathsf{E}[\boldsymbol{z}] = \mathbf{0}$, $\mathsf{E}[\boldsymbol{z}\boldsymbol{z}^\mathsf{H}] = \mathbf{I}$, and $\mathsf{E}[\boldsymbol{z}\boldsymbol{z}^\mathsf{T}] = \mathbf{0}$, where $\mathsf{E}[\cdot]$ denotes the expectation value. Note that $\sigma = 1/\sqrt{2}$ corresponds to the (pure) vacuum state, while larger values of $\sigma$ correspond to a (mixed) thermal state. The analogue of the squeezed state $\hat{\boldsymbol{b}}$ is then the random vector $\boldsymbol{b}$ defined by

$$\boldsymbol{b} = (\cosh \mathbf{R})\, \boldsymbol{a} + (\sinh \mathbf{R})\, \mathbf{Q}\, \boldsymbol{a}^*. \tag{3}$$

Transformations of this form appear in classical nonlinear mixing [20], so we may also view this as a classical model arising from nonlinear optics and reified (i.e., real, not virtual) vacuum modes. Our fundamental hypothesis is that $\boldsymbol{b}$ provides an accurate statistical representation of $\hat{\boldsymbol{b}}$ when applied to vacuum or thermal states.

Since $\boldsymbol{b}$ is a linear combination of complex Gaussian random variables, it, too, is a complex Gaussian random vector. As such, it is defined by a mean value, a covariance matrix, and a pseudo-covariance matrix. The mean is clearly zero, and the covariance is given by

$$\mathbf{\Gamma} = \mathsf{E}\left[bb^{\mathsf{H}}\right] = \sigma^2 \cosh(2\mathbf{R}) \,, \tag{4}$$

which is positive semi-definite. Unlike $a$, however, $b$ is not generally a proper random vector since the pseudo-covariance,

$$\mathbf{C} = \mathsf{E}\left[bb^{\mathsf{T}}\right] = \sigma^2 \left[(\cosh \mathbf{R})\mathbf{Q}^{\mathsf{T}} \sinh \mathbf{R}^{\mathsf{T}} + (\sinh \mathbf{R})\mathbf{Q} \cosh \mathbf{R}^{\mathsf{T}}\right] \,, \tag{5}$$

is not necessarily zero. In such cases, $b$ is said to be an *improper* complex Gaussian random vector [21,22].

Improper Gaussian random vectors arise in several signal and image processing applications. However, to date, they have received little attention within the physics community in relation to quantum optics and entanglement. Impropriety, a measure of the degree to which a random vector is improper, may be interpreted as a correlation between the real and imaginary parts of $b$. A popular measure of the *degree of impropriety* is the following:

$$\mathcal{I} = \frac{|\det \mathbf{C}|^2}{(\det \mathbf{\Gamma})^2} \,. \tag{6}$$

This definition is equivalent to others that have been proposed for characterizing improper random vectors [23]. It can be shown that $0 \leq \mathcal{I} \leq 1$ and, for proper random vectors, $\mathcal{I} = 0$. A random vector for which $\mathcal{I} = 1$ is considered *maximally* improper [24]. If $\mathbf{C}$ is singular but nonzero, then $b$ is improper but has zero impropriety. Note that the definition of impropriety may be applied to any complex random vector, whether it is Gaussian or not, provided the second moments are well defined.

As a statistical relationship between the real and imaginary parts of $b$ (or, equivalently, $b$ and $b^*$), we might expect impropriety to be fundamentally related to the commutation relations between the quadratures of $\hat{b}$ (or, equivalently, $\hat{b}$ and $\hat{b}^\dagger$). In quantum mechanics, this relationship is captured in such familiar quantities as the Mandel $Q_M$ parameter, which compares the mean and variance of the (normally ordered) number operator, and the squeezing parameter $S_\theta$, which measures the imbalance between the quadratures of a squeezed state [25–27]. For, say, a single squeezed vacuum mode, these nonclassicality parameters take on anomalous values precisely when the squeezing parameter $r$ is nonzero. This, as we now show, can be tied directly to impropriety.

Let us consider the special case $\mathbf{R} = r\mathbf{I}$, for $r \geq 0$. In this case, the covariance matrix is $\mathbf{\Gamma} = \sigma^2 \cosh(2r)$, $\mathbf{Q}$ is symmetric, and the pseudo-covariance matrix takes the simple form

$$\mathbf{C} = \sigma^2 \sinh(2r)\, \mathbf{Q} \,. \tag{7}$$

Since $\mathbf{Q}$ is unitary, the impropriety is found to be

$$\mathcal{I}(r) = \tanh(2r)^{2d} \,. \tag{8}$$

Note that $\mathcal{I}(r)$ grows monotonically with $r$ and is independent of both $\mathbf{Q}$ and $\sigma$. We further note that as the squeezing parameter increases (i.e., as $r \to \infty$), the degree of impropriety converges to unity, eventually approaching a state of maximum impropriety.

For this special case, it can furthermore be shown that the probability density function for $b$ is given by [28]

$$f(\boldsymbol{\beta}) \propto \exp\left[-\frac{1}{2}\begin{pmatrix}\boldsymbol{\beta}^{\mathsf{H}} & \boldsymbol{\beta}^{\mathsf{T}}\end{pmatrix}\begin{pmatrix}\mathbf{\Gamma} & \mathbf{C} \\ \mathbf{C}^* & \mathbf{\Gamma}^*\end{pmatrix}^{-1}\begin{pmatrix}\boldsymbol{\beta} \\ \boldsymbol{\beta}^*\end{pmatrix}\right] = \frac{\exp\left[-\|(\cosh r)\,\boldsymbol{\beta} - (\sinh r)\,\mathbf{Q}\,\boldsymbol{\beta}^*\|^2/\sigma^2\right]}{(\pi\sigma^2)^d} \,, \tag{9}$$

where $\boldsymbol{\beta} \in \mathbb{C}^d$. This matches precisely the Wigner function $W(\boldsymbol{\beta}, \boldsymbol{\beta}^*)$ for $\hat{b}$. This is of course unsurprising since the second moments $\mathsf{E}[b_i b_j^*]$ of $b$ match the symmetrized expectations $\langle(\hat{b}_i \hat{b}_j^\dagger + \hat{b}_j^\dagger \hat{b}_i)/2\rangle$ of $\hat{b}$.

Clearly, $f(\boldsymbol{\beta})$ and $W(\boldsymbol{\beta}, \boldsymbol{\beta}^*)$ are nonnegative (indeed, Gaussian) and, in this sense, classical, even though the $P$ function of $\hat{b}$ may not be. Nevertheless, the squeezed quantum state may still exhibit entanglement (i.e., nonseparability). The Peres–Horodecki criterion, extended to continuous variables, can be used to determine separability [29]. For the special case of, say, a symmetric two-mode Gaussian state with squeezing matrix

$$\boldsymbol{\xi} = re^{i\phi}\begin{pmatrix}0 & 1 \\ 1 & 0\end{pmatrix} \,, \tag{10}$$

the Peres–Horodecki criterion requires $\sigma^2 e^{-2r} < 1/2$ for entanglement. For the vacuum state ($\sigma^2 = 1/2$), we have entanglement for all $r > 0$, while a general thermal state will be entangled only if $r > (1/2)\log(2\sigma^2)$. For a general two-mode Gaussian quantum state, whether squeezed or not, propriety in the corresponding random vector (i.e., $\mathbf{C} = \mathbf{0}$) implies that the quantum state is separable [30]. The converse, as we have seen, need not be true. Thus, entanglement may imply impropriety, as it does in this case, but impropriety alone does not entail that the state is entangled.

## 3 Post-selected states

Let us consider again the general case $\boldsymbol{\xi} = \mathbf{RQ}$ and let $|\boldsymbol{\xi}\rangle = \hat{S}|\mathbf{0}\rangle$ denote the corresponding multi-modal squeezed vacuum state. If the degree of squeezing is small, as measured by some suitable norm on $\boldsymbol{\xi}$, we may approximate the squeezed state $|\boldsymbol{\xi}\rangle$ as follows:

$$|\boldsymbol{\xi}\rangle \approx |0,\ldots,0\rangle + \tfrac{1}{2}\sum_{ij}\xi_{ij}\hat{a}_i^\dagger \hat{a}_j^\dagger |0,\ldots,0\rangle. \tag{11}$$

Neglecting the vacuum state, the second term in this approximation represents a two-photon state of the form

$$|\psi\rangle \propto \tfrac{1}{2}\xi_{11}|2,0,\ldots,0\rangle + \xi_{12}|1,1,0,\ldots,0\rangle + \cdots + \xi_{1d}|1,0,\ldots,0,1\rangle + \cdots + \tfrac{1}{2}\xi_{dd}|0,\ldots,0,2\rangle. \tag{12}$$

Note that, although $|\boldsymbol{\xi}\rangle$ is Gaussian, the approximate post-selected state $|\psi\rangle$ generally is not.

Of particular interest will be squeezing matrices of the form

$$\boldsymbol{\xi} = r\begin{pmatrix} 0 & 0 & \alpha_1 & \alpha_2 \\ 0 & 0 & \alpha_3 & \alpha_4 \\ \alpha_1 & \alpha_3 & 0 & 0 \\ \alpha_2 & \alpha_4 & 0 & 0 \end{pmatrix}, \tag{13}$$

as these can be used to represent two-photon polarization states of the form

$$|\psi\rangle = \alpha_1|HH\rangle + \alpha_2|HV\rangle + \alpha_3|VH\rangle + \alpha_4|VV\rangle. \tag{14}$$

Here, the basis states $|HH\rangle, |HV\rangle,\ldots$ are used to represent the Fock states $|1,0,1,0\rangle, |1,0,0,1\rangle,\ldots$. The corresponding random vector $\boldsymbol{b}$ shall be denoted as $[b_{AH}, b_{AV}, b_{BH}, b_{BV}]^\mathsf{T}$, so the pair $(b_{AH}, b_{AV})$ may be associated with the first photon and $(b_{BH}, b_{BV})$ may be associated with the second photon.

Taking $\alpha_1 = \alpha_4 = 0$ and $\alpha_2 = -\alpha_3 = 1/\sqrt{2}$, for example, gives the Bell state $|\psi\rangle \propto |HV\rangle - |VH\rangle$. The polar decomposition in this case is $\boldsymbol{\xi} = \mathbf{RQ}$, with $\mathbf{R} = (r/\sqrt{2})\mathbf{I}$ and

$$\mathbf{Q} = \begin{pmatrix} 0 & 0 & 0 & 1 \\ 0 & 0 & -1 & 0 \\ 0 & -1 & 0 & 0 \\ 1 & 0 & 0 & 0 \end{pmatrix}. \tag{15}$$

As noted previously, the corresponding Gaussian random vector $\boldsymbol{b}$ is improper for all $r > 0$. For post-selected squeezed vacuum states, the corresponding quantum state $|\psi\rangle$ is, of course, maximally entangled. However, for general thermal states this need not be the case, even when $\boldsymbol{b}$ is improper.

By contrast, suppose $\alpha_1 = \alpha_2 = \alpha_3 = \alpha_4 = 1/2$, corresponding to a separable superposition of all four modes. In this case, $\boldsymbol{\xi} = \mathbf{UDV}^\mathsf{H}$, where $\mathbf{D} = \mathrm{diag}(r, r, 0, 0)$ and $\mathbf{U}, \mathbf{V}$ are unitary matrices taken to be

$$\mathbf{U} = \frac{1}{\sqrt{2}}\begin{pmatrix} 1 & 0 & -1 & 0 \\ 1 & 0 & 1 & 0 \\ 0 & -1 & 0 & -1 \\ 0 & -1 & 0 & 1 \end{pmatrix}, \tag{16}$$

$$\mathbf{V} = \frac{1}{\sqrt{2}}\begin{pmatrix} 0 & -1 & 0 & 1 \\ 0 & -1 & 0 & -1 \\ 1 & 0 & 1 & 0 \\ 1 & 0 & -1 & 0 \end{pmatrix}. \tag{17}$$

Using $\mathbf{R} = \mathbf{U}\mathbf{D}\mathbf{U}^H$ and $\mathbf{Q} = \mathbf{U}\mathbf{V}^H$, we find that the pseudo-covariance of $\boldsymbol{b}$ is nonzero but singular, since $\det \mathbf{R} = 0$ implies $\det \mathbf{C} = 0$. So, $\boldsymbol{b}$ is improper, but the degree if impropriety is zero, in keeping with the separability of $|\psi\rangle$.

## 4 Bell–CHSH inequality violations

To further verify entanglement, we considered computing a Bell statistic for the Clauser–Horne–Shimony–Holt (CHSH) inequality [17]. For the Bell statistic, we used the observables $\mathbf{A}_1 = \mathbf{Z}$, $\mathbf{A}_2 = \mathbf{X}$, $\mathbf{B}_1 = (\mathbf{Z} + \mathbf{X})/\sqrt{2}$, and $\mathbf{B}_2 = (\mathbf{Z} - \mathbf{X})/\sqrt{2}$, where $\mathbf{X}$ and $\mathbf{Z}$ are the Pauli $x$ and $z$ matrices. For quantum observables, the Bell statistic

$$S = |C_{11} + C_{12}| + |C_{21} - C_{22}|, \tag{18}$$

using $C_{ij} = \langle \psi | \mathbf{A}_i \otimes \mathbf{B}_j | \psi \rangle$, is $S = 2\sqrt{2}$.

For classical observables, a specific definition of measurement is needed to compute the correlations $C_{ij}$, and for this we used local amplitude threshold crossings as a model for single-photon detection [15,16]. In this approach, a detection of the modal component $b_i$ is said to occur when $|b_i| > \gamma$ for some fixed amplitude threshold $\gamma \geq 0$. In the analysis to follow, we take $\gamma = 1$ for all detectors.

Verification was done numerically as follows. First, a random sample of $N = 2^{20}$ realizations of the vacuum states $\boldsymbol{a} = [a_{AH}, a_{AV}, a_{BH}, a_{BV}]^T$ were generated, where the components of $\boldsymbol{a}$ are independent and identically distributed proper complex Gaussian random variables with zero mean and variance $\sigma^2 = 1/2$. Using Eqns. (3) and (15) with $\mathbf{R} = (r/\sqrt{2})\mathbf{I}$, we computed the squeezed modes $\boldsymbol{b}_A = [b_{AH}, b_{AV}]^T$ and $\boldsymbol{b}_B = [b_{BH}, b_{BV}]^T$ representing two entangled photons measured by Alice and Bob, respectively. Specifically, these are given by

$$\boldsymbol{b}_A = \begin{bmatrix} b_{AH}\cosh\left(r/\sqrt{2}\right) + b^*_{BV}\sinh\left(r/\sqrt{2}\right) \\ b_{AV}\cosh\left(r/\sqrt{2}\right) - b^*_{BH}\sinh\left(r/\sqrt{2}\right) \end{bmatrix}, \tag{19}$$

$$\boldsymbol{b}_B = \begin{bmatrix} b_{BH}\cosh\left(r/\sqrt{2}\right) - b^*_{AV}\sinh\left(r/\sqrt{2}\right) \\ b_{BV}\cosh\left(r/\sqrt{2}\right) + b^*_{AH}\sinh\left(r/\sqrt{2}\right) \end{bmatrix}. \tag{20}$$

In the simulation, $r$ was varied from 0 to 3. Note that, although $r \gg 1$ exceeds the validity condition of Eqn. (11), the equations for $\boldsymbol{b}_A$ and $\boldsymbol{b}_B$ remain valid.

For the random vectors $\boldsymbol{b}_A$ and $\boldsymbol{b}_B$, measurements were performed as follows. To measure $\mathbf{A}_1 = \mathbf{Z}$, Alice need only consider the components $b_{AH}$ and $b_{AV}$. Let $I_H$ denote the subsets of all realizations for which $|b_{AH}| > \gamma$, and let $I_V$ be the subset for which $|b_{AV}| > \gamma$. Events in $I_H$ correspond to an outcome of $+1$ for measuring $\mathbf{A}_1$, while events in $I_V$ correspond to an outcome of $-1$. By contrast, the complementary set $\bar{I}_H$ denotes the set of realizations for which an outcome of $+1$ did *not* occur, either because the outcome was $-1$ or because there was no detection observed by Alice.

To measure, say, $\mathbf{B}_2$, we first applied a unitary matrix $\mathbf{U}_-^\dagger$ to $\boldsymbol{b}_B$ to obtain $\boldsymbol{b}'_B = \mathbf{U}_-^\dagger \boldsymbol{b}_B = [b'_{BH}, b'_{BV}]^T$, where

$$\mathbf{U}_\pm = \begin{pmatrix} \cos(\pi/8) & \pm\sin(\pi/8) \\ \pm\sin(\pi/8) & -\cos(\pi/8) \end{pmatrix} \tag{21}$$

is such that $\mathbf{U}_+^\dagger \mathbf{B}_1 \mathbf{U}_+ = \mathbf{U}_-^\dagger \mathbf{B}_2 \mathbf{U}_- = \mathbf{Z}$ is diagonal. Let $J_H$ to denote the subset of all realizations for which $|b'_{BH}| > \gamma$, and, similarly, define $J_V$ to be those for which $|b'_{BV}| > \gamma$. Much as before, events in $J_H$ correspond to an outcome of $+1$ for measuring $\mathbf{B}_2$, while events in $J_V$ correspond to an outcome of $-1$. Note that measurements of $\mathbf{A}_1$ and $\mathbf{B}_1$ are each performed locally.

Now, the correlation $C_{12}$ between measurements of $\mathbf{A}_1$ and those of $\mathbf{B}_2$ was computed as follows:

$$C_{12} = (+1)(+1)\frac{\Pr[E_{HH}]}{\Pr[E]} + (+1)(-1)\frac{\Pr[E_{HV}]}{\Pr[E]} + (-1)(+1)\frac{\Pr[E_{VH}]}{\Pr[E]} + (-1)(-1)\frac{\Pr[E_{VV}]}{\Pr[E]}, \tag{22}$$

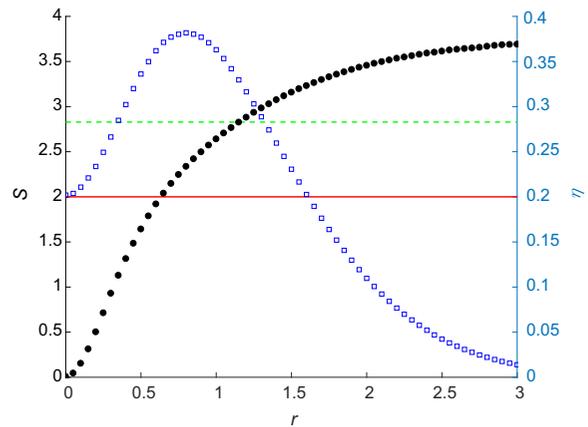

**Fig. 1** (color online) Plot of the Bell statistic $S$ (black dots) and coincident detection efficiency $\eta$ (blue squares) versus the squeezing parameter $r$ for an entangled Bell state. The solid red horizontal line is the classical bound of 2, while the dashed green horizontal line is the Tsirelson bound of $2\sqrt{2}$ where $\Pr[E_{HV}]/\Pr[E]$, say, is the probability of the event $E_{HV} = I_H \cap \bar{I}_V \cap \bar{J}_H \cap J_V$ conditioned on the set of coincident single-detection events $E = E_{HH} \cup E_{HV} \cup E_{VH} \cup E_{VV}$. Note that conditioning on coincident detection events corresponds to post-selecting for $|\psi\rangle$ upon preparing $|\xi\rangle$. (Higher order multi-photon terms are considered negligible if $r$ is small.) This post-selection is necessary to prepare the desired entangled state but introduces contextuality since the set $E$ will be different for each of the four measurement selections.

The other correlations are computed similarly, and doing so for all four combinations of observables results in the Bell statistic $S$. In Fig. 1, we have plotted $S$ as a function of $r$, the magnitude of the squeezing parameter. We see that for $r$ greater than about 0.5, we obtain a violation of the CHSH inequality $S \leq 2$. The ability to violate the CHSH inequality is, of course, a result of the post-selection performed on coincident single-detection events, which gives rise to contextuality and the detection loophole [31–33]. We furthermore note that for $r$ greater than about 1, the validity of Eqn. (11) is compromised and, so, we also get a violation of the Tsirelson bound of $2\sqrt{2}$ [34]. The upward trend continues monotonically towards an asymptote of 4, which is the algebraic upper bound on $S$.

Such large values for $S$ are not possible within quantum mechanics but do occur in post-quantum models such as Popescu–Rohrlich (PR) boxes [35]. Despite their unusual behavior, these models are not merely speculative but can be created artificially. For example, it has been observed that a PR box may be created simply through post-selection [36]. Indeed, an experimental realization of this for a three-photon state has already been performed [37]. A similar experiment using pairs of photons in entangled orbital angular momentum states was used to demonstrate near maximal violations [38]. Recently, a notional scheme for an optical realization of a PR box for two polarization photons has also been proposed [39]. The importance of the present work is to demonstrate how a physical model, such as we have described, can plausibly violate the fair-sampling hypothesis and result in extreme violations of the CHSH inequality, as have been observed experimentally.

High values of $S$ rely upon post-selection of rare events, which can lead to low coincident detection efficiencies. We define the efficiency, $\eta$, as the probability of a coincident detection, conditioned on a single detection for either measurement, and minimized over all measurements, as suggested in Ref. [33]. As shown in Fig. 1, the efficiency, $\eta$, is maximized near $r = 0.8$, which is the regime in which we get a CHSH violation. Much larger values of $r$, which are needed for more extreme violations, occur with vanishing probability. The maximum efficiency of about 38% is consistent with the detection loophole and comparable to what is observed for typical avalanche photodiodes [40].

## 5 Conclusion

In this paper, we have considered a classical model for certain multi-mode squeezed states in which the quantum mechanical annihilation operators are replaced with independent proper complex Gaussian random variables. The transformed random vector representing the squeezed state was found to be complex Gaussian distributed but not necessarily proper, due to the presence of a possibly nonzero pseudo-covariance matrix. Entangled quantum states were found to correspond to improper classical random vectors, but impropriety alone was not found to entail entanglement. The model was further examined by demonstrating violations of the Bell–CHSH inequality on post-selected coincident detections using an amplitude threshold crossing model of single-photon detection, with results that conform well with experimental observations for typical avalanche photodiode detectors.

**Acknowledgements** This work was funded by the Office of Naval Research under Grant no. N00014-18-1-2107.

**Data availability** The simulation tools used in this study are available from the corresponding author upon reasonable request.

**Declarations**

**Conflict of interest** The authors declare that they have no conflicts of interest.

**Publisher's Note** Springer Nature remains neutral with regard to jurisdictional claims in published maps and institutional affiliations.